\documentclass[aps,prb,twocolumn,showpacs,amsmath,amssymb,floatfix,superscriptaddress]{revtex4}
\usepackage{epsfig}
\usepackage{bm}
\usepackage{color}

\begin{document}

\title{
Thermal decay of the Coulomb blockade oscillations}

\author{Edvin G. Idrisov}
\affiliation{D\'epartement de Physique Th\'eorique, Universit\'e de Gen\`eve, CH-1211 Gen\`eve 4, Switzerland}

\author{Ivan P. Levkivskyi}
\affiliation{Theoretische Physik, ETH Zurich, CH-8093 Zurich, Switzerland}

\author{Eugene V. Sukhorukov}
\affiliation{D\'epartement de Physique Th\'eorique, Universit\'e de Gen\`eve, CH-1211 Gen\`eve 4, Switzerland}
\date{\today}

\begin{abstract}

We study transport properties and the charge quantization phenomenon in a small metallic island connected to the leads through two quantum point contacts (QPCs). The linear conductance is calculated perturbatively with respect to weak tunneling and weak backscattering at QPCs as a function of the temperature $T$ and gate voltage. The conductance shows Coulomb blockade  (CB) oscillations as a function of the gate voltage that decay with the temperature as a result of thermally activated fluctuations of the charge in the island. The regimes of quantum, $T \ll E_C$, and thermal, $T \gg E_C$, fluctuations are considered, where $E_C$ is the charging energy of an isolated island.  Our predictions for CB oscillations in the quantum regime coincide with previous findings in [A. Furusaki and K. A. Matveev, Phys. Rev. B {\bf 52}, 16676 (1995)]. In the thermal regime the visibility of Coulomb blockade oscillations decays with the temperature as  $\sqrt{T/E_C}\exp(-\pi^2T/E_C)$, where the exponential dependence originates from the thermal averaging over the instant charge fluctuations, while the prefactor has a quantum origin. This dependence  does not depend on the strength of couplings to the leads. The differential capacitance, calculated in the case of a single tunnel junction, shows the same exponential decay, however the prefactor is linear in the temperature. This difference can be attributed to the non-locality of the quantum effects.  Our results agree with the recent experiment [S. Jezouin {\em et al}., Nature {\bf 536}, 58 (2016)] in the whole range of the parameter $T/E_C$.
\end{abstract}

\maketitle

\section{Introduction}
\label{Sec:I} 

The transport of electrons through small mesoscopic conductors, such as metallic and semiconductor quantum dots,
has been extensively studied both experimentally and theoretically.\cite{Grabert} One of the most popular experimental systems in this field is the single electron transistor (SET), which consists of a quantum dot (typically, a small micrometer-scale metallic island) connected to two metallic leads 
by tunnel junctions and capacitively coupled to an additional gate electrode.\cite{Kastner} Such a  three-terminal device has been theoretically proposed by Averin and Likharev,\cite{Averin} and fabricated and characterized by Fulton and Dolan.\cite{Fulton} The most important characteristic of these devices, which sets them apart from the conventional field effect transistors, is that they can be switch between insulating and conducting state by adding as small amount of charge to the gate electrode as the charge of an electron $e$.

Such strong charge sensitivity of SETs is consequence of the fact that the charge of an isolated metallic island is quantized in units of the elementary charge $e$. Tunneling of an electron from a lead into the  island changes its charge by $e$ and increases the charging energy of the system by the amount of order  $E_C=e^2/2C$, where $C$ is a geometrical capacitance of the island. When the temperature is small, $T\ll E_C $,  tunneling is strongly  suppressed except at degenercy points between states, where the number of electrons in the island differs by one. Therefore, the conductance of the SET  exhibits a series of peaks as a function of the applied gate voltage. This well-known phenomenon, called the Coulomb blockade (CB) effect, is widely reviewed in the literature.\cite{Grabert,Kastner,Hawrylak,Ando} Recently, it has received renewing interest in the context of  the quantum information processing, where the semiconductor-metal hybrid devices serve as crucial elements for quantum computing, \cite{Larsen} and in the quest for topologically protected quantum bits. \cite{Albrecht}

\begin{figure}
\includegraphics[width=8cm]{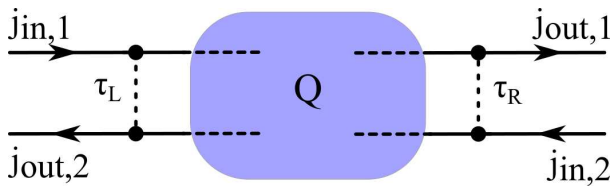}
\caption{\label{fig1} The experimental system in Ref.\ [\onlinecite{Pierre}] is schematically shown. In the integer quantum Hall regime, where only one spinless edge mode contributes to the transport,  a metallic island (containing the charge $Q$) is connnected to the leads by two QPCs, characterised by the backscattering amplitudes $\tau_L$ and $\tau_R$. The edge states are described by four bosonic operators, labeled by $\alpha=\text{in, out}$ and $j=1,2$. The average current $\langle I \rangle$ is calculated through the cross-section immediately to the right of the right QPC.}
\end{figure}

The growing interest to single electron phenomena stimulated further research and recently resulted in the publication of the thorough experimental study of the decay of CB oscillations  in mesoscopic circuits induced by thermal and quantum fluctuation of charge.\cite{Pierre} The experimental setup in Ref.\  [\onlinecite{Pierre}] is schematically shown in Fig.\ \ref{fig1}.
The experimentalists fabricated a hybrid SET based on a two-dimensional electron gas in the integer quantum Hall (QH) regime  by attaching  a central micrometer-scale metallic island to large electrodes  with the help of edge channels. Unlike in earlier experiments,\cite{Kouwenhoven,Staring,Van der Vaart} this new approach allows a pricese control of coupling of the island to the leads by using two QPC, which mix incoming and outgoing edge channels with the complex amplitudes $\tau_L$ and $\tau_R$. The metallic island has a negligible level spacing, which implies that it can be considered a reservoir for electron-hole excitations. The experimentalists measured the visibility 
of CB oscillations in the linear conductance of the SET as a function of the tunneling amplitudes $\tau_{L,R}$ and the temperature $T$ in the whole range of the parameters. The purpose of the present paper is to present the theory and interpretation of the observed effects in the regimes of weak tunneling and weak backscattering at QPCs, which are theoretically accessible by using the tunneling Hamiltonian appoach.

Between a large number of previous theoretical works on the CB effect (see, e.g., Refs.\ [\onlinecite{Korshunov,Schon, Flensberg, Nazarov,Chouvaev,Golubev,Titov,Burmistrov}]) the earlier theory [\onlinecite{Furusaki}] of Furusaki and Matveev deserves a special attention in the present context, because they have addressed the CB oscillations with the help of the model described above (see the Fig.\ \ref{fig1}). Using the standard bosonization technique they have made quantitative predictions for the visibility of CB oscillations in the quantum regime $T\ll E_C$ in the case of strong tunneling, where one or both QPCs are close to perfect transmission. Here we extend these predictions to arbitrary temperatures by using an alternative approach to bosonization,\cite{Slobodeniuk,Sukhorukov1} which is based on the scattering theory for bosons as well as the Langevin equation method. Our quantative predictions fully agree with the experimental data in the whole range of temperatures. In the thermal regime, $T\gg E_C$, the amplitude of CB oscillations in the conductance scales as  $\delta G \propto \sqrt{T/E_C}\exp(-\pi^2 T/E_C)$, independently of the geometry of the system. Here, the exponential function results from the averaging of CB oscillations over the classical thermal fluctuations of the charge in the island, while the power-law prefactor has a quantum origin. Interestingly, the CB oscillations in the differential capacitance of the island acquire the linear in $T$ prefactor, which points to the non-universality of this contribution and to its quantum origin.

The rest of the paper is organized as follows. In Sec.\ \ref{Sec:II},  we introduce the model of the system, starting with the Hamiltonian of all the constituting parts and the bosonization prescription, and formulate the method of quantum Langevin equations.  In Sec.\ \ref{Sec:III}, we calculate the linear conductance and the visibility of CB oscillations for the case of the symmetric setup, i.e., for weak backscattering at both QPCs, using the perturbation approach in backscattering amplitudes. In Sec.\ \ref{Sec:IV}, we develop the perturbation theory for the case of the asymmetric setup with weak tunneling at one of the two QPCs and calculate the conductance and the visibility in this regime. Sec.\ \ref{Sec:V} is devoted to the derivation of the differential capacitance. We present our conclusions and the discussion in Sec.\ \ref{Sec:VI}. Details of calculations are given in appendices. Throughout the paper, we set $e=\hbar=k_B=1$.

\section{Theoretical model}
\label{Sec:II}
\subsection{Hamiltonian}
We start by introducing the Hamiltonian of the experimental system in the Ref.\ [\onlinecite{Pierre}], see Fig.\ \ref{fig1}.  The relevant energy scales in this experiment are sufficiently small compared to the Fermi energy, $\epsilon_F$, which suggests using the effective low-energy theory of QH edge states. \cite{Wen} The advantage of this approach is that it allows to take into account Coulomb interaction at the metallic island in a straightforward way. \cite{Slobodeniuk} According to the effective theory, edge states can be described as collective fluctuations of the charge densities $\rho_{\alpha j}(x)$, where indexes label the number of the channel,   $j=1,2$, and the state, $\alpha=\text{in,out}$. The charge density operators are expressed in terms of bosonic fields $\phi_{\alpha j}(x)$, namely, $\rho_{\alpha j}(x)=(1/2\pi)\partial_x\phi_{\alpha j}(x)$.
These free bosonic fields satisfy canonical commutation relations
\begin{equation}
\label{Canonical commutation relations for free bosonic fields}
[\partial_x \phi_{\alpha j}(x),\phi_{\beta k}(y)]=(-1)^{\alpha} 2\pi i\delta_{\alpha\beta}\delta_{jk}\delta(x-y),
\end{equation}
where the sign determines the propagation direction of the edge states.

The total Hamiltonian includes three terms  
\begin{equation}
\label{Total Hamiltonian}
H=H_0+H_{\text{int}}+H_{\text{T}}.
\end{equation}
Here
\begin{equation}
\label{Kinetic term of Hamiltonian}
H_0=\frac{v_F}{4\pi}\sum_{\substack{\alpha=\text{in,out} \\ j=1,2}}\int dx
[\partial_x\phi_{\alpha j}(x)]^2
\end{equation}
is the free part of the total Hamiltonian, and the Fermi velocity, $v_F$, is the same
for each edge channel.
The second term describes Coulomb interaction at the metallic island:
\begin{equation}
\label{Coulomb interaction term of Hamiltonian}
H_{\text{int}}=\frac{(Q-Q_0)^2}{2C},
\end{equation}
where 
\begin{equation}
\label{Operator of the total charge on the metallic grain}
\begin{split}
& Q = \frac{1}{2\pi} \int_{0}^{\infty} dx \left[\partial_x \phi_{\text{in} 1}(x)+\partial_x \phi_{\text{out} 2}(x)\right] \\
& + \frac{1}{2\pi} \int_{-\infty}^{0} dx \left[\partial_x \phi_{\text{in} 2}(x)+\partial_x \phi_{\text{out} 1}(x)\right]
\end{split}
\end{equation}
is the operator of the total charge on the metallic island, $C$ being its geometrical capacitance, and the parameter $Q_0$ is proportional to the gate voltage and the capacitance $C_g$ between the island and the gate, namely $Q_0=C_g V_g$. Thus the  role of the gate is to control the average charge of the metallic island. 

The last term describes  backscattering of electrons at the left and right QPC, 
\begin{equation}
\label{Tunneling Hamiltonian}
\begin{split}
& H_{\text{T}}= A_{L}+A_{R}+ \text{H.c.}, \\ 
& A_{L}=\frac{\tau_{L}}{a}e^{i\phi_{\text{in} 1}(0)-i\phi_{\text{out} 2}(0)}, \\
& A_{R}=\frac{\tau_{R}}{a}e^{i\phi_{\text{out} 1}(0)-i\phi_{\text{in} 2}(0)},
\end{split}
\end{equation}
where $\tau_i$ are the tunneling coupling constants, and $a$ is the ultraviolet cut-off. \cite{Slobodeniuk, Sukhorukov, Levkivskyi, Slobodeniuk1} Note, that here we set the distance between the grain and QPC's to zero, because in the experiment it is much shorter than the characteristic wavelength of excitations, $\lambda \sim v_{F}/T$.

The Hamiltonian (\ref{Total Hamiltonian}) gives the complete description of our system. We note, that
the part $H_0+H_{\text{int}}$ is quadratic in bosonic operators, thus the dynamics
associated with this Hamiltonian can be accounted exactly by solving linear equations of motion. We follow the Refs.\ [\onlinecite{Slobodeniuk,Sukhorukov1}] and complement these equations  with the boundary conditions in terms of dissipative currents originating from the reservoirs and the metallic island. Fluctuations of these currents are  Gaussian, therefore, they can be considered Gaussian sources in the so-arising quantum Langevin equations. In contrast, the tunneling term (\ref{Tunneling Hamiltonian}) is non-linear in bosonic operators. Therefore, in Sec.\ \ref{Sec:III} and Sec.\ \ref{Sec:IV} we develop the perturbation theory to leading order in backscattering and tunneling  amplitudes.

\subsection{Quantum Langevin equations}

As the first step, we consider the currents in  the system shown in  Fig.\ \ref{fig1}, when QPCs are fully open (symmetric setup), and write the equations of motion generated by the part $H_0+H_{\text{int}}$ of the total Hamiltomian in the following form:\cite{Slobodeniuk}

\begin{equation}
\label{Langevin equations}
\begin{split}
& \frac{dQ(t)}{dt}= \sum_{\alpha=1,2} j_{\text{in} \alpha}(t)-\sum_{\alpha=1,2} j_{\text{out} \alpha}(t), \\
& j_{\text{out} \alpha}(t)=\frac{Q(t)-Q_0}{R_q C}+j_{\alpha}^{s}(t),
\end{split}
\end{equation} 
where $R_q=2\pi$ is a quantum resistance. Here, 
the first equation expresses the conservation of charge. The second line is the Langevin equations, which have the following physical meaning. The outgoing currents acquire two contributions:
the first one is the current induced by the time dependent potential $Q(t)/C$, and the second, $j_{\alpha}^{s}$, are the Langevin current sources originating from the metallic island.

Now, we are ready to calculate the current $\langle I \rangle_0$ through the system in absence of backscattering, i.e., in the case of fully open QPCs. We define it as the average current in the cross-section immediately after the right QPC,
\begin{equation}
\label{Bare current}
\langle I \rangle_0 = \langle j_{\text{out} 1} \rangle -\langle j_{\text{in} 2}\rangle,
\end{equation}
Assuming, that the first channel is voltage biased with $\Delta\mu$, while the second is grounded, we set $\langle j_{2\text{in}}\rangle=0$. Averaging the Eqs.\ (\ref{Langevin equations}) and taking into account that the average incoming current is equal to the average outgoing current, we obtain
\begin{equation}
\label{Averaging of Langevin equations}
\begin{split}
& \langle j_{\text{out} 1} \rangle+\langle j_{\text{out} 2}\rangle=\Delta \mu / R_q, \\
& \langle j_{\text{out} 1} \rangle=\langle j_{\text{out} 2}\rangle=\frac{\langle Q \rangle - Q_0}{R_q C}.
\end{split}
\end{equation}
Next, we use these equations to calculate the average value for the total charge stored in the  metallic island, $\langle Q \rangle=Q_0+\Delta \mu C/2$, and the average current through the system, $\langle I\rangle_0 =\langle j_{\text{out} 1}\rangle=\Delta \mu/2 R_q$.  
Therefore, in the absence of backscattering the conductance acquires the value
\begin{equation}
\label{Bare conductance}
G_0 = \frac{1}{2}G_q,
\end{equation}
where $G_q=1/R_q=1/2\pi$ is the conductance quantum, i.e., the system behaves as two quantum resistances connected in series.

\begin{figure}
\includegraphics[width=8cm]{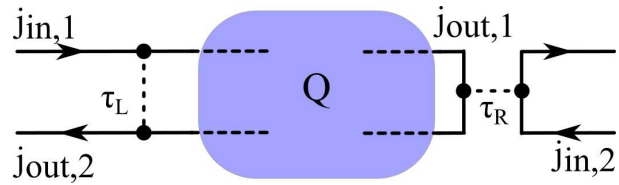}
\caption{\label{fig2. Theoretical description of the experimental setup in case of asymmetric barriers.} The asymmetric setup, where the left QPC is almost fully open, while the right QPC is weakly transmitting,  is schematically shown.  We investigate CB oscillations to leading order in the backscattering amplitude $\tau_L$ and tunneling amplitude $\tau_R$. }
\end{figure}

According to the equations (\ref{Langevin equations}), the fluctuating currents satisfy the following equations:
\begin{equation}
\label{Langevin equations for the fluctuations of currents}
\begin{split}
& \frac{d}{dt}  \delta Q(t)= \sum_{\alpha=1,2} \delta j_{\text{in} \alpha}(t)-\sum_{\alpha=1,2} \delta j_{\text{out} \alpha}(t), \\
& \delta j_{\text{out} \alpha}(t)=\frac{\delta Q(t)}{R_q C}+\delta j_{\alpha}^{s}(t). 
\end{split}
\end{equation} 
These equations can be easily solved in the frequency representation, and for the symmetric setup we obtain:
\begin{equation}
\label{Scattering matrix in frequency representation. Symmetric barriers}
\begin{pmatrix}
  \delta j_{\text{out} 1}(\omega) \\ \delta j_{\text{out} 2}(\omega)
\end{pmatrix}
=
\begin{pmatrix}
  a(\omega) & b(\omega) & -b(\omega) & -b(\omega) \\ b(\omega) & a(\omega) & -b(\omega) & -b(\omega)
\end{pmatrix}
\begin{pmatrix}
  \delta j^{s}_{1}(\omega) \\ \delta j^{s}_{2}(\omega) \\ \delta j_{\text{in} 1}(\omega) \\ \delta j_{\text{in} 2}(\omega)
\end{pmatrix},
\end{equation}
where $a(\omega)=(i\omega R_q C-1)/(i\omega R_q C-2)$ and $b(\omega)=1/(i\omega R_q C-2)$ are the scattering coefficients. 
Next, in order to address a particular experimental situation,\cite{Pierre} we also consider  asymmetric setup, i.e., where for instance the left QPC  is fully open, while the right one is fully closed (see Fig.~\ref{fig2. Theoretical description of the experimental setup in case of asymmetric barriers.}).
We skip the details of the calculations, which are analogous to those in the case of the symmetric setup, and present the result:
 \begin{equation}
\label{Scattering matrix in frequency representation. Asymmetric barriers}
\begin{split}
& \delta j_{\text{out} 1}(\omega)=\delta j^{s}_1(\omega)+\frac{1}{i\omega R_q C - 1}\left[\delta j^{s}_2(\omega)-\delta j_{\text{in} 1}(\omega)\right], \\
& \delta j_{\text{out} 2}(\omega)=\frac{i\omega R_q C}{i\omega R_q C - 1}\delta j^{s}_1(\omega) - \frac{1}{i\omega R_q C -1}\delta j_{\text{in} 1}(\omega).
\end{split}
\end{equation}

Finally, the two-point correlation functions of the incoming currents and Langevin sources are given by the equilibrium spectral function\cite{Lifshitz}
\begin{equation}
\label{Equilibrium spectral function}
\langle \delta j^s_{\alpha}(\omega)\delta j^s_{\beta}(\omega^{\prime})\rangle = \langle \delta j_{\text{in} \alpha}(\omega)\delta j_{\text{in} \beta}(\omega^{\prime})\rangle = \delta_{\alpha \beta}\delta(\omega + \omega^{\prime})S(\omega),
\end{equation}
where $S(\omega)=2\pi G_q \omega/(1-e^{-\omega/T})$.  With the help of Eqs.\ (\ref{Scattering matrix in frequency representation. Symmetric barriers}), (\ref{Scattering matrix in frequency representation. Asymmetric barriers}) and (\ref{Equilibrium spectral function}), one obtains two-point correlators of the currents at QPCs.
In the following sections we use these correlators to derive the corrections to the bare conductance perturbatively in the cases of the symmetric and asymmetric setup. Below we will use that $R_q=2\pi$ and $G_q=1/2\pi$ explicitly. These quantities are easily restored in final results for conductance.

\section{Symmetric setup}
\label{Sec:III}
We consider the regime of low bias, $\Delta\mu\ll T$, and evaluate the linear conductance through the metallic island in the symmetric setup shown in the Fig.\ \ref{fig1}:
\begin{equation}
\label{Linear conductance}
G=\left. \frac{d\langle I\rangle}{d\Delta \mu}\right|_{\Delta \mu=0}.
\end{equation}
In the interaction representation the average current is given by the expression
\begin{equation}
\label{Average current. Symmetric barriers}
\langle I \rangle = \text{Tr}\left[\rho_0 U^{\dagger}(t,-\infty)I(t)U(t,-\infty)\right],
\end{equation}   
where
\begin{equation}
\label{Evolutional operator. Symmetric barriers}
U(t_1,t_2) = \hat{\text{T}}\text{exp}\left[-i\int_{t_2}^{t_1} dt H_{\text{T}}(t)\right]
\end{equation} 
is the evolution operator. The current operator is defined in the cross-section immediately after the right QPC, 
\begin{equation}
\label{Current operator. Symmetric barriers}
I(t)=-\frac{1}{2\pi}\partial_t\left[\phi_{\text{out} 1}(x_r,t)-\phi_{\text{in} 2}(x_r,t)\right],
\end{equation}
and $\rho_0 \propto \text{exp}\left[-(H_0+H_{\text{int}})/T\right]$ is the equilibrium density matrix.

We evaluate the average current perturbatively expanding the evolution operator to the lowest order in backscattering  amplitudes,
\begin{equation}
\label{Total current. Symmetric barriers}
\langle I \rangle = \langle I \rangle_0 +\delta \langle I \rangle_0,
\end{equation}
where the  non-perturbed current $\langle I \rangle_0=\Delta \mu/ 2 R_q$ was derived in the previous section, and the second term can be written as (for details, see the Appendix \ref{Sec:A})
\begin{align}
\label{Kubo formula for current correction. Four terms. Symmetric barriers}
\delta \langle I \rangle_0 = \sum_{l,l^{\prime}}I_{ll^{\prime}}, \quad I_{ll^{\prime}}= -\frac{1}{2}\int dt \langle \left[A^{\dagger}_{l}(t),A_{l^{\prime}}(0)\right]\rangle_0,
\end{align}
where the average is taken with respect to the equilibrium non-perturbed state with the density matrix $\rho_0$. Here, $I_{\text{dir}}=I_{LL}+I_{RR}$ is the direct contribution to the current, while $I_{\text{osc}}=I_{LR}+I_{RL}$ is the oscillating part. The later one oscillates as a function of the parameter $Q_0$, proportional to the gate voltage, as $\cos(2\pi Q_0)$, which is the manifestation of the CB effect. Finally,  the prefactor 1/2 in this expression comes from the fact, that the metallic island mixes the edge channels equally even in the absence of backscattering.

\subsection{Quantum regime}

We evaluate the correction $\delta G$ to the linear conductance by using the Kubo formula (\ref{Kubo formula for current correction. Four terms. Symmetric barriers}) and split  it in a direct and oscillating part: $\delta G=\delta \langle I \rangle_0/\Delta\mu=G_{\text{dir}}+G_{\text{osc}}$.  In the low-temperature, quantum regime, $T\ll E_C $ we obtain (see Appendix \ref{Sec:B} for details) 
\begin{equation}
\label{Incoherent term for conductance. Quantum regime. Symmetric barriers}
G_{\text{dir}}= -\frac{\mid \tau_L \mid^2 + \mid \tau_R \mid^2 }{v^2_{F}}\frac{e^{\bold{\gamma}}E_C}{8\pi^2T},
\end{equation}
where $\bold{\gamma} \approx 0.5772$ is the Euler's constant. For the  oscillating contribution in this regime we obtain the expression 
\begin{equation}
\label{Coherent term for conductance. Quantum regime. Symmetric barriers}
G_{\text{osc}} = -\frac{2\mid \tau_L \mid \mid \tau_R \mid}{v^2_{F}}\frac{e^{\bold{\gamma}}E_C}{8\pi^2T}\cos(2\pi Q_0).
\end{equation}
Taking into account that $G_q=1/2\pi$ and combining the equations (\ref{Bare conductance}), (\ref{Incoherent term for conductance. Quantum regime. Symmetric barriers}), and (\ref{Coherent term for conductance. Quantum regime. Symmetric barriers}), we present the total linear  conductance in the following form
\begin{equation}
\label{Conductance. Quantum regime. Symmetric barriers}
G=\frac{G_q}{2}\left(1-\frac{\Gamma(Q_0)}{T}\right) ,
\end{equation}
where 
\begin{equation}
\Gamma(Q_0)= \frac{e^{\bold{\gamma}}E_C}{2\pi v^2_{F}}\left[|\tau_L|^2 + |\tau_R|^2 + 2|\tau_L||\tau_R|\cos(2\pi Q_0)\right].
\end{equation}
It is important to mention, that in 
 Eq.\ (\ref{Conductance. Quantum regime. Symmetric barriers}) we kept only leading order terms in the parameter $E_C/T$, and according to our perturbation approach in weak backscattering, $T \gg \Gamma(Q_0)$. In this limit, our results fully agree with the earlier theory of Furusaki and Matveev. \cite{Furusaki} Interestingly, the same expressions were derived for the conductance  of a 1D system with two defects.\cite{Nagaosa} 

\subsection{Thermal regime}

In the thermal regime,  $T\gg E_C $, we derive (see Appendix \ref{Sec:B} for details) the following expressions for direct
\begin{equation}
\label{Incoherent term for conductance. Thermal regime. Symmetric barriers}
G_{\text{dir}}=-\frac{1}{8\pi}\frac{\mid \tau_L \mid^2 + \mid \tau_R\mid^2}{v^2_{F}}
\end{equation}
and the oscillating term
\begin{equation}
\label{Coherent term for conductance. Thermal regime. Symmetric barriers}
G_{\text{osc}}=-\frac{|\tau_L||\tau_R|}{2v^2_{F}}\sqrt{\frac{\pi T}{E_C}}\exp\left[-\frac{\pi^2 T}{E_C}\right]\cos(2\pi Q_0).
\end{equation}
Taking into account that $G_q=1/2\pi$ and combining the Eqs.\ (\ref{Bare conductance}), (\ref{Incoherent term for conductance. Thermal regime. Symmetric barriers}), and (\ref{Coherent term for conductance. Thermal regime. Symmetric barriers}), we obtain the following expression for the total linear conductance: 
\begin{equation}
\label{Conductance.Thermal regime.Symmetric barriers}
G=\frac{G_q}{2}\left[1-\frac{|\tau_L|^2+|\tau_R|^2+2|\tau_L||\tau_R|F(T)\cos(2\pi Q_0)}{2v^2_{F}}\right],
\end{equation}
where the temperature dependent factor is given by
\begin{equation}
\label{F}
F(T)=2\pi\sqrt{\frac{\pi T}{E_C}}\exp\left[-\frac{\pi^2 T}{E_C}\right].
\end{equation}
In contrast to the quantum regime, here the temperature influences only  CB oscillations.

We note, that these results have an interesting interpretation. While at high temperatures the bare currents
at the left and write QPC can be considered being independent 1D currents, at low temperatures, $T\ll E_C$, (and consequently, at low frequencies) the metallic island splits them equally. This leads to the phenomenon of charge fractionalization (half of electron is reflected by the island), and thus reduces the exponent of the power-law correlators in (\ref{Kubo formula for current correction. Four terms. Symmetric barriers}) by the factor of 2. As a result, the direct conductance in (\ref{Incoherent term for conductance. Quantum regime. Symmetric barriers}) acquires the singular temperature dependence $1/T$.

\subsection{Visibility}

\begin{figure}
\includegraphics[width=8cm]{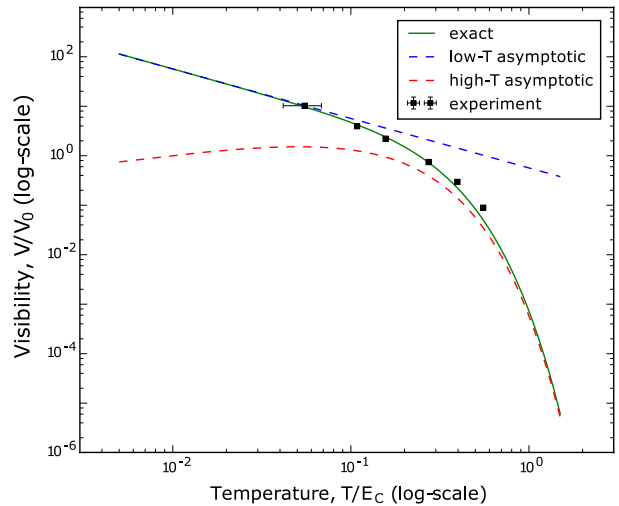}
\caption{\label{fig3.Temperature dependence of visibility in case of symmetric barriers.} The exact temperature dependence of the visibility (green line) is shown for the case of the symmetric setup together with the asymptotic low- and high-temperature solutions (dashed lines) and experimental results (black squares). The visibility is normalized to the value $V_0=|\tau_L||\tau_R|/v_F^2$ in order to get rid of the non-universal backscattering amplitudes.  The experimental results are obtained for various transmissions at the left QPC\cite{Pierre}:  $G_L/G_q= 0.983$, $0.974$, $0.974$, $0.75$, $0.75$, $0.75$ for $T/E_C = {0.055,  0.108,  0.157,  0.273,  0.397, 0.553}$ respectively. The transmission at the right QPC, $G_R/G_q$, is kept very close to $1$.}
\end{figure}

The strength of the CB oscillations is described by the visibility function:
\begin{equation}
\label{Visibility}
V = \frac{G_{\text{max}}-G_{\text{min}}}{G_{\text{max}}+G_{\text{min}}}.
\end{equation} 
In the quantum regime the visibility is given by 
\begin{equation}
\label{Visibility. Quantum regime. Symmetric case}
V = \frac{|\tau_L||\tau_R|e^{\bold{\gamma}}E_C}{\pi v^2_{F}T},\quad T\ll E_c,
\end{equation}
and according to the perturbation approach in weak backscattering, $V\ll 1$.
In the thermal regime we obtain the following result
\begin{equation}
\label{Visibility. Thermal regime. Symmetric case}
V= \frac{2|\tau_L||\tau_R|}{v^2_{F}}\pi\sqrt{\frac{\pi T}{E_C}}\exp\left[-\frac{\pi^2 T}{E_C}\right], \quad T\gg E_C.
\end{equation}
The exact dependence of the visibility on the temperature in  whole range of temperatures, from quantum to thermal regime, is found by calculating time integrals in the Appendix  \ref{Sec:B} numerically. The results are presented in Fig.\ \ref{fig3.Temperature dependence of visibility in case of symmetric barriers.} together with asymptotic solutions 
(\ref{Visibility. Quantum regime. Symmetric case}) and (\ref{Visibility. Thermal regime. Symmetric case}), and the results of the experiment [\onlinecite{Pierre}]. It is worth mentioning a good agreement with the experiment. 
 
\section{Asymmetric setup}
\label{Sec:IV}
In this section we study the case of the asymmetric setup, namely, we assume that the right QPC is almost closed (pinched-off), while the left one is almost open (see Fig.\ \ref{fig2. Theoretical description of the experimental setup in case of asymmetric barriers.}). To find the conductance of the system we apply the perturbation expansion in two steps: we first calculate the linear conductance perturbatively in the tunneling at the right QPC,  and then consider corrections to it taking into account weak backscattering at  the left QPC.

 The electron current is defined as a rate of change of the electron number, $N_R$, in the right arm, namely
\begin{equation}
\label{Current operator. Assymetric case}
I_R = i\left[H_T,N_R\right]=i(A_R-A_R^\dagger) .
\end{equation}
We evaluate the average current to the lowest order in the tunneling amplitude $\tau_R$:
\begin{equation}
\label{Kubo formula. Assymetric case}
\langle I_R \rangle = \int dt \langle [A^{\dagger}_R(t),A_R(0)]\rangle .
\end{equation}
Here, the average is taken with respect to the density matrix 
\begin{equation}
\label{Density matrix. Assymetric case}
\rho=U(0,-\infty)\rho_0U^\dagger(0,-\infty),
\end{equation} 
perturbed by weak backscattering at the left QPC. To the lowest order in the backscattering amplitude $\tau_L$, the evolution operator is given by the expression
\begin{equation}
\label{Evolution operator. Assymetric case}
U(t_1,t_2)=1-i\int_{t_2}^{t_1} dt ( A_L+A_L^\dagger).
\end{equation}

Using the Eq.\  (\ref{Kubo formula. Assymetric case}), we present the linear conductance  $G=\langle I_R\rangle/\Delta\mu$ in the following form:
\begin{equation}
\label{G. Assymetric case}
G=\frac{i|\tau_R|^2}{a^2} \int dt\, t \left[K_0(t)G_1(t)-K^{\ast}_0(t)G_2(t)\right],
\end{equation}
where
\begin{equation}
\label{Correlation function of free fermions. Assymetric case}
K_0(t)=\frac{-iT}{2v_{F}\sinh[\pi T(t-i0)]}
\end{equation}
is the two-point correlation function of free fermions in the right arm of the system in Fig.\ \ref{fig2. Theoretical description of the experimental setup in case of asymmetric barriers.}.
The correlation functions 
\begin{eqnarray}
G_1(t)&=&\langle U^\dagger(t,-\infty)e^{-i\phi_{\text{out} 1}(t)}\nonumber\\  &{}&\qquad\quad\times
U(t,0)e^{i\phi_{\text{out} 1}(0)}U(0,-\infty)\rangle_0,
\label{G_1. Assymetric case}
\end{eqnarray} 
\begin{eqnarray}
G_2(t)&=&\langle U^\dagger(0, -\infty)e^{i\phi_{\text{out} 1}(0)}\nonumber \\
&{}&\qquad\quad\times U(0,t)e^{-i\phi_{\text{out} 1}(t)}U(t,-\infty)\rangle_0
\label{G_2. Assymetric case}
\end{eqnarray}
are calculated perturbatively to the lowest order in the backscattering amplitude, $\tau_L$,  i.e., by using the Eq.\ (\ref{Evolution operator. Assymetric case}).
After expanding  Eqs.\ (\ref{G_1. Assymetric case}) and (\ref{G_2. Assymetric case}) and substituting them into the Eq.\ (\ref{G. Assymetric case}), one can split the conductance $G$ in the sum of two terms
\begin{equation}
\label{Conductance. Asymmetric case}
G=G_{\text{dir}} + G_{\text{osc}}.
\end{equation}
Here, the term $G_{\text{dir}}$ is proportional to $|\tau_R|^2$, and does not depend on the gate parameter $Q_0$. In contrast, the contribution $G_{\text{osc}}$ is proportional to $|\tau_L||\tau_R|^2$ and demonstrates CB oscillations. Detailed calculations of these two terms are presented in the Appendix \ref{Sec:C}.
  
\subsection{Quantum regime}

Repeating the same steps as in the Sec.\ \ref{Sec:III},  we  obtain the following expression for the direct and oscillating contributions to the conductance in the limit of low temperatures, $T\ll E_C$, 
(see Appendix \ref{Sec:C})
\begin{equation}
\label{Incoherent term for conductance. Quantum regime. Asymmetric barriers}
G_{\text{dir}}(T) = \frac{2\pi^4T^2G_{R}}{3e^{2\bold{\gamma}}E^2_C},
\end{equation}
where $G_R=|\tau_R|^2/2\pi v^2_{F}$ is the bare conductance of the right QPC. The oscillating contribution  acquires the form
\begin{equation}
\label{Coherent term for conductance. Quantum regime. Asymmetric barriers}
G_{\text{osc}}=-\frac{2\pi^4 T^2G_{R}}{3e^{2\bold{\gamma}}E^2_C}\xi \frac{|\tau_L|}{v_{F}}\cos(2\pi Q_0).
\end{equation}
Substituting Eqs.\ (\ref{Incoherent term for conductance. Quantum regime. Asymmetric barriers}) and (\ref{Coherent term for conductance. Quantum regime. Asymmetric barriers}) into  the Eq.\ (\ref{Conductance. Asymmetric case}), we arrive at the following expression for the total linear conductance
\begin{equation}
\label{Conductance.Quantum regime. Assymmetric barriers}
G=\frac{2\pi^4 T^2G_R}{3e^{2\bold{\gamma}}E^2_C}\left[1-\xi \frac{|\tau_L|}{v_{F}}\cos(2\pi Q_0)\right],
\end{equation}
where $\xi=4e^{\bold{\gamma}}$ is the dimensionless numeric constant.\cite{Furusaki} Note, that even in the absance of backscattering at the left QPC the conductance scales as $T^2$.

\subsection{Thermal regime}

In this case, $T\gg E_C$, we get the following expressions for the direct and oscillating contributions to the total conductance (details of calculations are presented in the Appendix \ref{Sec:C}): 
\begin{equation}
\label{Incoherent term for conductance.Thermal regime.Asymmetric barriers}
G_{\text{dir}}=\frac{|\tau_R|^2}{2\pi v^2_{F}}=G_R,
\end{equation}
and 
\begin{equation}
\label{Coherent term for conductance.Thermal regime.Asymmetric barriers}
G_{\text{osc}}=-G_R\frac{|\tau_L|}{v_{F}}M(T)\cos(2\pi Q_0),
\end{equation}
where the temperature dependent factor is given by
\begin{equation}
\label{M}
M(T)=2\pi \sqrt{\frac{\pi T}{E_C}}\exp\left[-\frac{\pi^2 T}{E_C}\right].
\end{equation}
Combining two terms, we present the total conductance as
\begin{equation}
\label{Conductance.Thermal regime.Asymmetric barriers}
G=G_R\left[1-\frac{|\tau_L|}{v_{F}}M(T)\cos(2\pi Q_0)\right].
\end{equation}
Note, that the temperature affects only the oscillating contribution, in contrast to the quantum regime. 

\subsection{Visibility}

As it follows from the Eq.\ (\ref{Conductance.Quantum regime. Assymmetric barriers}), the visibility in the quantum regime is constant:
\begin{equation}
\label{Visibility. Quantum regime. Assymmetric case}
V= \xi \frac{|\tau_L|}{v_{F}},\quad T\ll E_C.
\end{equation} In the thermal regime [see Eq.\ (\ref{Conductance.Thermal regime.Asymmetric barriers})], the visibility acquires the following form 
\begin{equation}
\label{Visibility. Thermal regime. Assymmetric case}
V =\frac{2\pi |\tau_L|}{v_{F}}\sqrt{\frac{\pi T}{E_C}}\exp\left[-\frac{\pi^2 T}{E_C}\right], \quad T\gg E_C.
\end{equation}
The results of the exact numerical calculation of time integrals for the visibility in the Appendix \ref{Sec:C} are shown in Fig.\ \ref{fig4.Temperature dependence of visibility in case of asymmetric barriers.} together with the asymptotic
 forms and the experimental data [\onlinecite{Pierre}]. Our results agree quite well with the experiment. 
\begin{figure}
\includegraphics[width=8cm]{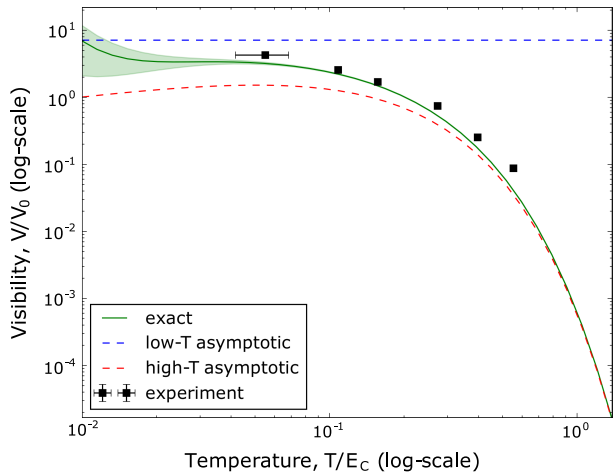}
\caption{\label{fig4.Temperature dependence of visibility in case of asymmetric barriers.} The exact temperature dependence of the visibility (green line) is shown for the case of the asymmetric setup together with the asymptotic low- and high-temperature solutions (dashed lines) and experimental results (black squares). The visibility is normalized to the value $V_0= |\tau_L|/v_F$ in order to get rid of the non-universal backscattering amplitude.  The experimental results are obtained for transmission at the left QPC\cite{Pierre}:  $G_L/G_q= 0.075$ for $T/E_C = {0.055,  0.108,  0.157,  0.273,  0.397, 0.553}$ and the transmission at the right QPC, $G_R/G_q$, is kept very close to $1$. The green shadowed region shows the error bars of the numerical evaluation. }
\end{figure}

\section{Differential capacitance}
\label{Sec:V}
Without loss of generality, we consider here an asymmetric case, namely, we assume that the right QPC is pinched off with $\tau_R=0$ (see Fig.\ref{fig2. Theoretical description of the experimental setup in case of asymmetric barriers.}), while the left QPC is almost open with the weak backscattering amplitude $\tau_L$. In this case, the CB effect manifests itself in  the oscillations of the equilibrium characteristics of the system, such as its ground state energy or the average charge of the island. The single-electron capacitance spectroscopy of quantum dots \cite{Ashoori1, Ashoori2} can be used to measure experimentally the differential capacitance between the gate and the lead. The differential capacitance is defined, up to a non-universal constant prefactor, as
\begin{equation}
\label{Differencial capacitance}
C_{\text{diff}} = -\frac{\partial^2 F }{\partial Q^2_0},
\end{equation}
where $F$ is the free energy of the system,
\begin{equation}
\label{Free Energy}
F = -T\log(Z),
\end{equation}
and $Z= Tr(e^{-H/T})$ is the partition function. 

The fully open island does not exhibit the QB effect, therefore we evaluate the partition function perturbatively in the backscattering amplitude $\tau_L$ and concentrate on the correction to the capacitance. We write
\begin{equation}
Z=Z_0+\delta Z,
\end{equation}
where $Z_0=\text{Tr}\left[e^{-(H_0+H_{\text{int}})/T}\right]$ is the bare part, while
the correction is given by
\begin{equation}
\delta Z = -\text{Tr}\left[ e^{-(H_0+H_{\text{int}})/T}\int_0^{1/T}d\tau H_{\text{T}}(\tau)\right], 
\end{equation}
where $\tau$ is the imaginary time.
Consequently, the free energy can be rewritten as
\begin{equation}
F=F_0+\delta F,
\end{equation}
where $F_0=-T\log(Z_0)$, and the correction has the following form
\begin{equation}
\label{Correction to free energy}
\delta F = - T\frac{\delta Z}{Z_0}=\text{Tr}(\rho_0 H_{\text{T}}),
\end{equation}
where $H_{\text{T}}=A_L+A_L^\dagger$.

Next, we write the right hand side of Eq.\ (\ref{Correction to free energy}) in bosonic operators using the definition 
(\ref{Tunneling Hamiltonian}), and evaluate the correlation
function $
\langle e^{i\phi_{\text{out}2}(0)}e^{-i\phi_{\text{in}1}(0)}\rangle_0$
by using the scattering matrix (\ref{Scattering matrix in frequency representation. Asymmetric barriers}) for bosons in the asymmetric setup and the Gaussian character of the theory. Thus, we obtain
\begin{equation}
\label{Correction to free energy. Exact expression1} 
\delta F = \frac{2|\tau_L|}{a}\cos(2\pi Q_0)e^{y(T)},
\end{equation}
where 
\begin{equation}
y(T)=-\int \frac{dx x }{1+x^2}\frac{1}{1-e^{-x/2\pi TC}}.
\end{equation}
We evaluating $y(T)$ in the limit of high temperatures, $T\gg E_C$:
\begin{equation}
y(T)=\log(a/4\pi^2 v_{F} C)-2\pi^2 TC + \log(4\pi^2 TC),
\end{equation}
and find the correction to the free energy
\begin{equation}
\label{Correction to free energy. Exact expression2}
\delta F = \frac{2|\tau_L|}{v_{F}}Te^{-\pi^2 T/E_C}\cos(2\pi Q_0).
\end{equation}
Consequently, the correction to the differential capacitance in the thermal regime is given by
\begin{equation}
\label{C}
\delta C_{\text{diff}}=8\pi^2\frac{|\tau_L|}{v_{F}}T\exp\left[-\frac{\pi^2 T}{E_C}\right]\cos(2\pi Q_0).
\end{equation}
Interestingly, the differential capacitance contains the same exponential factor as in the linear conductance for both setups. This  can be easily explained  by the fact that the exponential dependence on the temperature simply follows from averaging CB oscillations over instant fluctuations of the charge in the island, which are distributed with the equilibrium Gibbs weights $\rho_G \propto \exp[-Q^2 E_C/T]$. However, the power-law prefactor in the differential capacitance is different from the one in the linear conductances [see Eqs.\ (\ref{F}) and (\ref{M})], which can be understood taking into account its quantum character, since one of the contributions to it comes from high-energy modes.

\section{Conclusion}
\label{Sec:VI}

The charge of an isolated metallic system  is quantized in units of the elementary electron charge. This phenomenon manifests itself in oscillations of the conductance, if a system is attached to metallic leads, and in the oscillations of the differential capacitance. The degree of the charge quantization is described by the dimensionless visibility $V$ of such oscillations. In the recent experiment conducted in the group of F.\ Pierre, \cite{Pierre} the transport through a small metallic island connected to two leads by QPCs (as schematically shown in Fig.\ \ref{fig1}) has been thoroughly studied in the regime, where the QPCs are pinched off to allow only one mode to partially propagate. In such systems quantum and thermal fluctuations of the current at QPC gradually reduce the quantization of charge as the tunneling coupling strength at QPCs is increased, and the CB oscillations vanish for fully open contacts. The work [\onlinecite{Pierre}] has reported measurements of the visibility of CB oscillations as a function of temperature as well as of the tunneling coupling strength at QPCs.

Motivated by this experiment, we have developed a quantitative theory of the linear conductance and differential capacitance of the metallic island in the whole range of temperatures, from below to above the charging energy $E_C$, in the cases of symmetric and asymmetric tunneling coupling at QPCs. To do so, we have use the quantum Langevin equation approach to account for the dynamics and fluctuations of the collective charge fluctuations in a fully open system, and the tunneling Hamiltonian approach to account for weak backscattering and weak tunneling of electrons at QPCs. \cite{Slobodeniuk, Sukhorukov}  This method is based on the fact that the electrical circuit elements typically create only Gaussian fluctuations and therefore the fully open system in the bosonic picture remains Gaussian, because the interaction part of Hamiltonian is quadratic in bosonic fields. 

We have found that in the low-temperature quantum regime, $T\ll E_C$, the temperature dependence of the  linear conductance  coincides with the results of Furusaki and Matveev \cite{Furusaki} in both cases of symmetric and asymmetric setup. For instance, in the symmetric case (see Fig.\ \ref{fig1}) the direct and oscillating part of the weak backscattering contribution to the  conductance of the system both acquire the power-law temperature dependence
 $\delta G \propto E_C/T$. On the other hand, in the case of asymmetric setup (see Fig.\ \ref{fig2. Theoretical description of the experimental setup in case of asymmetric barriers.}) 
the total conductance including the oscillating correction
acquires the temperature dependence  $G \propto (T/E_C)^2$, thus the visibility of CB oscillations stays constant. Different  power-law scaling of the conductance with the temperature depending on the geometry of the system indicates that an important role is played by the character of mixing of the collective modes in the metallic island. For instance, in the case of the symmetric setup the metallic island splits equally incoming currents, and thus the scaling $E_C/T$ originates from the charge fractionalization induced by such a current splitter, which is known to affect the power-law exponents. 

In the high-temperature limit, $T\gg E_C$, the temperature dependence of the oscillating correction to the conductance is entirely different: it is given by the product of a power-law prefactor and exponentially decaying function, $\delta G \propto \sqrt{T/E_C}\exp(-\pi^2 T/E_C)$, both for the symmetric and asymmetric setup. In order to compare our theory to the experiment [\onlinecite{Pierre}], we have found  the visibility of CB oscillations exactly in the entire range of temperatures by calculating time integrals numerically. Our results are shown in Figs.\ \ref{fig3.Temperature dependence of visibility in case of symmetric barriers.} and \ref{fig4.Temperature dependence of visibility in case of asymmetric barriers.} together with the asymptotic forms and the results of measurements. This comparison shows a good agreement of our theory with the experiment.

Finally, we have investigated CB oscillations in a partially closed system, where the right QPC is disconnected from the circuit ($\tau_L=0$, see Fig.\ \ref{fig2. Theoretical description of the experimental setup in case of asymmetric barriers.}), by evaluating the differential capacitance of the metallic island. Interestingly, the oscillating correction to the differential conductance in the high-temperature limit  acquires the same exponential decay with the temperature as the one for the linear conductance. Such universality can be explained by the fact that this effect is completely classical and originates from the thermal averaging of the CB oscillations over instant configurations of the charge in the island with the equlibrium Gibbs weights. On the contrary, the power-law prefactor is different [see Eq.\ (\ref{C})], which can be attributed to its quantum origin.

\acknowledgments
We are grateful to F. Pierre for clarifying the experimental data and L.I. Glazman for fruitful discussions. This work was supported by the Swiss National Science Foundation.
 
\appendix
\begin{widetext} 
\section{Derivation of the Kubo formula for symmetric setup}
\label{Sec:A}
In this section we derive the Eq.\ (\ref{Kubo formula for current correction. Four terms. Symmetric barriers}).  The operator of outgoing current to the right from the right QPC at point $x_r>0$  (see Fig.\ \ref{fig1}) expressed in bosonic fields reads
\begin{equation}
I(t)=-\frac{1}{2\pi}\partial_t\left[\phi_{\text{out} 1}(x_r,t)-\phi_{\text{in} 2}(x_r,t)\right].
\end{equation}  
Then, the average current is given by the following expression
\begin{equation}
\label{Current appendixA}
\langle I \rangle = \langle U^{\dagger}(t,-\infty)I(t)U(t,-\infty)\rangle_0,
\end{equation}
where the average is taken with respect to the unperturbed state $\rho_0 \propto \text{exp}\left[-(H_0+H_{\text{int}})/T\right]$, and  the evolution operator in interaction picture is expanded to second order in tunneling Hamiltonian Eq.\ (\ref{Tunneling Hamiltonian}):
\begin{equation}
\label{Evolution operator appendixA}
U(t,-\infty) = 1-i\int_{-\infty}^{t}dt_1H_{\text{T}}(t_1)-\int_{-\infty}^t dt_1 \int_{-\infty}^{t_1} dt_2 H_{\text{T}}(t_1)H_{\text{T}}(t_2).
\end{equation}
By substituting (\ref{Evolution operator appendixA}) into (\ref{Current appendixA}) and doing simple algebra, one arrives at the expression (\ref{Total current. Symmetric barriers}). 

The first term in (\ref{Total current. Symmetric barriers}) has been evaluated in Sec.\ \ref{Sec:II} with the result (\ref{Bare current}). Here, we concentrate on the second term in (\ref{Total current. Symmetric barriers}) and rewrite it in the following  form
\begin{equation}
\label{Correction to current. Symmetric case. Appendix}
\delta \langle I \rangle = -\int_{-\infty}^{t}dt_1 \int_{-\infty}^{t_1} dt_2 \left \langle \left[[I(t),H_{\text{T}}(t_1)], H_{\text{T}}(t_2)\right]\right \rangle_0.
\end{equation}
The integrand contains four non-zero terms, 
\begin{equation}
\label{Integrand}
\langle\left[\left[I(t),H_\text{T}(t_1)\right],H_\text{T}(t_2)\right]\rangle_0=
\Pi_1+\Pi_2+\text{c.c.},
\end{equation}
where the first two terms read
\begin{equation}
\begin{split} 
& \Pi_1=\left \langle \left[[I(t),A_L(t_1)],A^{\dagger}(t_2)\right]\right \rangle_0 ,\\
& \Pi_2=\left \langle \left[[I(t),A_R(t_1)],A^{\dagger}(t_2)\right]\right \rangle_0.
\end{split}
\end{equation}

Next, we concentrate on the first term $\Pi_1$ (the second term $\Pi_2$ can be evaluated in the same way) and simplify the commutator in $\Pi_1$
\begin{equation}
\label{Inner commutator}
\left[I(t),A_L(t_1)\right] \propto \left[\partial_t\phi_{\text{out} 1}(x_r,t)-\partial_t\phi_{\text{in} 2}(x_r,t),e^{i\left[\phi_{\text{in}1}(t_1)-\phi_{\text{out}2}(t_1)\right]}\right]
\end{equation} 
by using the following properties of operators. If an operator  $D=[B,C]$ satisfies $[B,D]=[C,D]=0$, then the following relations hold: $[B,\exp(C)]=[B,C]\exp(C)$ and $\exp(B)\exp(C)=\exp(B+C)\exp\left([B,C]/2\right)$. Thus, we obtain
\begin{equation}
\label{First term of integrand appendixA}
\Pi_1=R(t-t_1)K_L(t_1-t_2),
\end{equation}
where $K_L(t_1-t_2)=\langle \left[A_L(t_1),A^{\dagger}(t_2)\right] \rangle_0$ and $R(t-t_1)=-\frac{1}{2\pi}\left[\partial_t\phi_{\text{out} 1}(x_r,t)-\partial_t\phi_{\text{in} 2}(x_r,t),i\left[\phi_{\text{in}1}(t_1)-\phi_{\text{out}2}(t_1)\right]\right]$
Then, using the Langevin equations (\ref{Scattering matrix in frequency representation. Symmetric barriers}) and the commutator of incoming currents $[j_{\text{in}\alpha}(\omega),j_{\text{in}\beta}(\omega^{\prime})]=\omega \delta(\omega+\omega^{\prime})\delta_{\alpha \beta}$, we obtain the relation 
\begin{equation}
R(t-t_1)=\frac{1}{4\pi}\int \frac{d\omega \omega^2}{\omega^2+\eta^2}\left(\frac{e^{i\omega x_r/v_F}}{i\pi \omega C-1}-\frac{e^{-i\omega x_r/v_F}}{i\pi \omega C+1}\right)e^{-i\omega(t-t_1)}.
\end{equation}
It can be shown similarly, that 
\begin{equation}
\label{Second term of integrand appendixA}
\Pi_2=F(t-t_1)K_R(t_1-t_2),
\end{equation}
where $K_R(t_1-t_2)=\langle \left[A_R(t_1),A^{\dagger}(t_2)\right] \rangle_0$, and $
F(t-t_1)=-\delta(t-x_r/v_F-t_1)-\delta(t+x_r/v_F-t_1)-R(t-t_1)$.

Finally, substituting expressions (\ref{First term of integrand appendixA}) and (\ref{Second term of integrand appendixA}) into Eq.\ (\ref{Integrand}), and then into Eq.\ (\ref{Correction to current. Symmetric case. Appendix}), we arrive at the following expression
\begin{equation}
\label{Appendix1}
\delta\langle I\rangle=-\frac{1}{4\pi}\int d\omega\left(\frac{K(\omega)}{i\omega +\alpha}+\frac{K^{\ast}(\omega)}{-i\omega +\alpha}\right),
\end{equation}
where $\alpha \to +0$, and $K(\omega)=K_L(\omega)+K_R(\omega)$. We use the property $K(\omega)=K^{\ast}(\omega)$, which follows from $K^{\ast}(t)=\langle[A(t),A^{\dagger}(0)]\rangle^{\ast}=\langle[A(0),A^{\dagger}(t)]\rangle=K(-t)$. Thus, only $\omega =0$ contributes in the integral, and we obtain
\begin{equation}
\delta \langle I\rangle=-\frac{1}{2}\int dt \langle [A^{\dagger}(t),A(0)]\rangle_0,
\end{equation}
and eventually, the  Eq.\ (\ref{Kubo formula for current correction. Four terms. Symmetric barriers}) in the main text.

\section{Perturbation theory for symmetric setup}
\label{Sec:B}
In this appendix we derive the leading order corrections to the conductance of the system in the symmetric setup.
We use the Gaussian character of the theory to present the average of four vertex operators in the Eq.\ (\ref{Kubo formula for current correction. Four terms. Symmetric barriers}) in the following form
\begin{equation}
\langle e^{i\lambda_1 \phi_1}e^{i\lambda_2 \phi_2}e^{i\lambda_3 \phi_3}e^{i\lambda_4 \phi_4}\rangle_0=\exp\left(-\frac{1}{2}\sum_{i=1}^{4}\lambda^2_i \langle \phi^2_i \rangle_0 -\sum_{i<j}^4 \lambda_i \lambda_j \langle \phi_i \phi_j\rangle_0\right),\quad 
\lambda_i = \pm 1,
\end{equation}
where the average is taken with respect to the equilibrium density matrix $\rho_0 \propto \exp[-(H_0+H_{\text{int}})/T]$. 
We evaluate two-point correlation functions on the right hand side of this expression using quantum Langevin equation approach (\ref{Scattering matrix in frequency representation. Symmetric barriers}). According to Eq.\ (\ref{Kubo formula for current correction. Four terms. Symmetric barriers}), the correction to the average current consists of the direct and oscillating part, $
\delta \langle I \rangle_0 = I_{\text{dir}} + I_{\text{osc}} $.
Here we consider them separately. 

The direct term has the following form
\begin{equation}
\label{AppDir}
I_{\text{dir}} = -\frac{\mid \tau_L \mid^2 + \mid \tau_R \mid^2 }{2}\int dt e^{i\Delta \mu t/2}\left[K_{\text{ff}}(t)e^{-\beta(t)} - \text{c.c.} \right],
\end{equation}
where $\beta(t)$ is given below, and  
\begin{equation}
\label{Kff}
K_{\text{ff}}(t) = \frac{1}{a^2}\exp\left(2\int \frac{d\omega \omega}{\omega^2 + \eta^2}\frac{e^{-i\omega t}-1}{1-e^{-\omega/T}}\right)=\frac{-T^2}{4v^2_{F}\sinh^2[\pi T(t-i0)]}
\end{equation}
is the square of the free fermion correlation function. 
The oscillating term is given by the following expression 
\begin{equation}
\label{AppInt}
I_{\text{osc}} = -2\mid \tau_L \mid \mid \tau_R \mid \text{Re}\left[\frac{e^{-2 \pi iQ_0 + \alpha_S(T) }}{2a^2}\int dt e^{i\Delta \mu t/2}\left(e^{\beta(t)} - \text{c.c.}\right)\right].
\end{equation} 
Here
\begin{equation}
\alpha_{S}(T) = -2 \int dx \frac{x}{1+x^2}\frac{1}{1-e^{-x/\pi TC}} = 2\bold{\gamma} + 2\log\left(\frac{a}{2\pi^2 v_{F}C}\right) + 2\pi^2 TC + 2\log(2\pi^2 TC) + 2\Psi\left(1/2\pi^2 TC \right),
\end{equation}
$\Psi(x)=d \log(\Gamma(x))/dx$ is the digamma function of real variable $x$,  $\Gamma(x)$ is the gamma function, and $\beta(t) = \beta_1(t) + \text{Re}\beta_2(t) + i\text{Im}\beta_2(t)$.

Below, we need the expressions for $\beta_1(t)$ and $\beta_2(t)$ only for positive times, $t > 0$, where they take the following form:
\begin{equation}
\beta_1(t) = -2\int_0^{\infty} \frac{d\omega}{\omega^2+\eta^2}\frac{\omega}{1+(\pi \omega C)^2}\frac{\cos(\omega t)-1}{1-e^{\omega/T}},
\end{equation}
\begin{equation}
\text{Re}\beta_2(t)=\int_0^{\infty} d\omega \frac{\omega}{\omega^2+\eta^2}\frac{\cos(\omega t)-1}{1+(\pi \omega C)^2} = -\bold{\gamma}-\log(t/\pi C) + \frac{e^{-t/\pi C}}{2}\text{Ei}(t/\pi C) + \frac{e^{t/\pi C}}{2}\text{Ei}(-t/\pi C),
\end{equation}
\begin{equation}
 \text{Im}\beta_2(t) = -\int_0^{\infty} d\omega \frac{\omega}{\omega^2+\eta^2}\frac{\sin(\omega t)}{1+(\pi \omega C)^2} =
    -\frac{\pi}{2}\left(1-e^{-t/\pi C}\right).
\end{equation}
Here, $\text{Ei}(x)=- \int_{-x}^{\infty} dy e^{-y}/y$ is the exponential integral for real non zero values of $x$. Below, we use derived in this appendix integral representations for the current, take the derivative with respect to $\Delta\mu$ to obtaine the conductance, and find analytic expressions for the thermal, $T\gg E_C$, and quantum, $T\ll E_C$, regimes.

\subsection{Quantum regime}
We first consider the direct term (\ref{AppDir}). The contribution from the poles to the time integral at small times $ t \sim 1/\epsilon_F $ is a constant of temperature, while the dominant contribution to  (\ref{AppDir}) scales as $E_C/ T$. Therefore, we neglect the contributions from poles and present the  direct part of the  conductance as an integral over positive times, because the integrand is an even function of variable $t$ in this case:
\begin{equation}
\label{ApGdir}
G_{\text{dir}}= -\frac{\mid \tau_L \mid^2 + \mid \tau_R \mid^2 }{v^2_{F}}\int_0^{\infty} dt t\frac{T^2}{4\sinh^2(\pi Tt)}e^{-\beta_1(t)-\text{Re}\beta_2(t)}\sin\left[\frac{\pi}{2}(1-e^{-t/\pi C})\right]
\end{equation}
Next, taking into account that $T\ll E_C$, we set $C=0$ in $\beta_1(t)$   and obtain  $\beta_1(t)=-\log\left(\frac{\sinh(\pi Tt)}{\pi Tt}\right)$. Then, the main contribution to integral for $G_{\text{dir}}$ comes from large times, namely $t/\pi C \gg 1$.
In this case, we can simplify $\text{Re}\beta_2(t)=-\bold{\gamma}-\log(t/\pi C)$ and use $\sin\left[\frac{\pi}{2}(1-e^{-t/\pi C})\right] \sim 1$. After substitution these expressions in Eq.\ (\ref{ApGdir}), we obtain
\begin{equation}
\label{ApDir1}
G_{\text{dir}}=-\frac{\mid \tau_L \mid^2 + \mid \tau_R \mid^2 }{v^2_{F}} \frac{e^{\bold{\gamma}}}{4\pi^4 TC}\int_0^{\infty} dx \frac{x}{\sinh(x)} = -\frac{\mid \tau_L \mid^2 + \mid \tau_R \mid^2 }{v^2_{F}}\frac{e^{\bold{\gamma}}E_C}{8\pi^2T}.
\end{equation} 

In the oscillating term, at small temperatures $T\ll E_C$, the factor $\alpha_S(T)$ simplifies as $\alpha_S(T) \approx 2\bold{\gamma} + 2\log(a/2\pi^2 v_{F}C)$. Then, using the same arguments as for the direct term, we obtain 
\begin{equation}
\label{ApInt1}
G_{\text{osc}} = -\frac{2\mid \tau_L \mid \mid \tau_R \mid}{v^2_{F}}\cos(2\pi Q_0)\frac{e^{\bold{\gamma}}}{4\pi^4TC}\int_0^{\infty}dx\frac{x}{\sinh(x)} =  -\frac{2\mid \tau_L \mid \mid \tau_R \mid}{v^2_{F}}\frac{e^{\bold{\gamma}}E_C}{8\pi^2T}\cos(2\pi Q_0)
\end{equation}
Combining (\ref{ApDir1}), (\ref{ApInt1}), and (\ref{Bare conductance}), we obtain the Eq.\ (\ref{Conductance. Quantum regime. Symmetric barriers}) in the main text.

\subsection{Thermal regime}
In this regime, $T\gg E_C$, the prefactor $\alpha_S(T)$ has the following form
\begin{equation}
\alpha_S(T) \approx 2\log(a/2\pi^2 v_{F}C)-2\pi^2 TC + 2\log(2\pi^2 TC).
\end{equation}
We evaluate $\beta(t)$ by expanding the distribution function $1/(1-e^{-\omega/T})$ at large temperatures: 
\begin{equation}
\beta(t)=\int d\omega \frac{\omega}{\omega^2+\eta^2}\frac{e^{-i\omega t}-1}{1+(\pi \omega C)^2}\left[\frac{T}{\omega}+\frac{1}{2}\right] = \pi^2 TC \left(1-e^{-\mid t/\pi C \mid}-\mid t/\pi C \mid\right)-\frac{i\pi}{2}\left(1-e^{-\mid t/\pi C \mid}\right)\text{sign}(t),
\end{equation}
and then expanding  the right hand side of this equation  in series of $t/\pi C \ll 1$, taking into account the fact that the main contribution to the time integral comes from small times,  and that $T\gg E_C$. Thus, we obtain the following expression
\begin{equation}
\beta(t) = -\frac{Tt^2}{2C}-\frac{it}{2C}.
\end{equation}
Now, we are ready to write the correction for the conductance. 

The direct term reads
\begin{equation}
G_{\text{dir}}=\frac{\mid \tau_L\mid^2 + \mid \tau_R\mid^2}{16v^2_{F}}\int dt it T^2e^{Tt^2/2C}\left(\frac{e^{i t/2C}}{\sinh^2[\pi T(t-i0)]} - \text{c.c.}\right) .
\end{equation}
In this integral the main contribution comes from small times $t \sim 1/\epsilon_F$,  and we can neglect exponential factors in the integrand. Introducing the dimensionless variable $x=\pi Tt$, we write 
\begin{equation}
\label{Incoherent term thermal appendixC}
G_{\text{dir}}=\frac{\mid \tau_L\mid^2 + \mid \tau_R\mid^2}{16\pi^2v^2_{F}}\int dx ix \left(\frac{1}{\sinh^2[x-i0]} - \frac{1}{\sinh^2[x+i0]}\right) = -\frac{1}{8\pi}\frac{\mid \tau_L \mid^2 + \mid \tau_R\mid^2}{v^2_{F}}
\end{equation}
Next, we take derivative with respect to $\Delta\mu$ in Eq.\ (\ref{AppInt}) to get the following expression for the oscillating part:
\begin{equation}
G_{\text{osc}}=-\frac{4\mid \tau_L \mid \mid \tau_R \mid}{v^2_{F}}T^2C^2 e^{-2\pi^2 TC}\cos(2\pi Q_0)\int dy y \sin y\exp(-2TCy^2),
\end{equation}
where we introduced the dimensionless variable $y=t/2C$. 
Evaluating this integral, we obtain
\begin{equation}
\label{Coherent term thermal appendixC}
G_{\text{osc}}=-\frac{|\tau_L||\tau_R|}{2v^2_{F}}\sqrt{\frac{\pi T}{E_C}}\exp\left[-\frac{\pi^2 T}{E_C}\right]\cos(2\pi Q_0).
\end{equation} 
Here, we neglected the factor $\exp(-1/4\pi^2 TC)$, because $T\ll E_C$. Combining (\ref{Incoherent term thermal appendixC}), (\ref{Coherent term thermal appendixC}), and (\ref{Bare conductance}), we arrive at the expression for conductance (\ref{Conductance.Thermal regime.Symmetric barriers}) in the main text.   

\section{Perturbation theory for asymmetric setup}
\label{Sec:C}
In this appendix, we use Eqs.\  (\ref{G. Assymetric case}-\ref{Conductance. Asymmetric case})  to derive the leading order contribution to the conductance in the case of the asymmetric setup. We recall, that the direct contribution $G_{\text{dir}}$ is proportional to $|\tau_R|^2$, while the oscillating term $G_{\text{osc}}$ comes as a correction proportional to $|\tau_L||\tau_R|^2$ due to weak backscattering at the left QPC, and thus acquires oscillations as a function of the parameter $Q_0$. We derive analytical expressions in the quantum, $T\ll E_C$, and classical, $T\gg E_C$, regimes. 

\subsection{Quantum regime}
We first concentrate on the direct contribution.
We use the solution  Eq.\ (\ref{Scattering matrix in frequency representation. Asymmetric barriers}) of quantum Langevin equations for the asymmetric setup to evaluate bosonic correlators that arise in the perturbative expansion for the conductance and arrive at  the following expression
\begin{equation}
\label{G0freeAs}
G_{\text{dir}}=-\frac{|\tau_R|^2}{4v^2_{F}} \int dt it\left(\frac{T^2}{\sinh^2[\pi T(t-i0)]}e^{2\beta(t)}-\text{c.c.}\right) ,
\end{equation}
where $\beta(t)$ is given by the same expressions as in Sec.\ \ref{Sec:B} with $C$ replaced by $2C$, simply because in the asymmetric setup the conductance is twice as small. 
The expression (\ref{G0freeAs}) holds for arbitrary temperatures and serves as a starting point for evaluating low- and high-temperature limits. 
The contribution from the poles at small times $ t \sim 1/\epsilon_F $ takes the constant value $|\tau_R|^2/2\pi v^2_{F}$. Using the same arguments as in case of the symmetric setup (see Appendix \ref{Sec:B}), we obtain the expression for the direct part in the form
\begin{equation}
G_{\text{dir}}= \frac{|\tau_R|^2}{2\pi v^2_{F}}-\frac{|\tau_R|^2}{v^2_{F}} \int_0^{\infty} dt \frac{\pi^2T^4 t^3}{\sinh^4(\pi Tt)}e^{2\text{Re}\beta_2(t)}\sin[\pi(1-e^{-t/2\pi C})] .
\end{equation}
In the low-temperature limit, $T\ll E_C$, the main contribution to the integral in $G_{\text{dir}}$ comes from times much smaller then the inverse temperature, $t \ll 1/\pi T$. 
Thus, we can expand the temperature dependent part of the integrand in small $Tt$:
\begin{equation}
\label{G0Assym}
G_{\text{dir}}= \frac{|\tau_R|^2}{2\pi v^2_{F}}-\frac{4|\tau_R|^2}{v^2_{F}} \int_0^{\infty} dt\pi^2T^4 t^3\left[\frac{1}{\pi^4T^4t^4}-\frac{2}{3\pi^2T^2t^2}\right]e^{2\text{Re}\beta_2(t)}\sin[\pi(1-e^{-t/2\pi C})] .
\end{equation}

From the last equation it is obvious that $G_{\text{dir}}$ consists of a constant part and a temperature dependent part. The former can be presented as
\begin{equation}
\label{G0const}
G_{\text{dir}}(T=0) = \frac{|\tau_R|^2}{2\pi v^2_{F}} - \frac{|\tau_R|^2}{2\pi v^2_{F}}C_1.
\end{equation} 
It turnes out, that  the constant $C_1$ is exactly equal to 1, and $G_{\text{dir}}(T=0)$ vanishes. However, the easiest way to prove this fact is to rewrite the equation (\ref{G0freeAs}) for $T=0$ in the form
\begin{equation}
\label{integral quantum regime. AppendixB}
G_{\text{dir}}(T=0) \propto \int ds s\left[\frac{e^{M(s)}}{(s-i0)^2}-\frac{e^{M^{\ast}(s)}}{(s+i0)^2}\right] ,
\end{equation} 
where $M(s)= 2 \int_0^{\infty} \frac{dx (e^{-ix s}-1)}{x(1+x^2)}$,  and $s$ and $x$ are dimensionless variables. Here we note, that $M(s)$ is an analytical function in the low half plane. Moreover, it behaves as $-2\log|s|$ at $\text{Im}(s) \to -\infty$. Therefore, in the first term of the integral (\ref{integral quantum regime. AppendixB}) the contour may be deformed to $\text{Im}(s) \to -\infty$. Next, at $\text{Im}(s) \to -\infty$ the integrand behaves as $1/|s|^3$, and thus the integral vanishes. The same arguments are applied to the second term, where $M^{\ast}(s)$ is analytical in the upper half plane. Thus, we have proven, that $G_{\text{dir}}(T=0)=0$.

Now, we consider the temperature dependent term. To integrate  over $t$ in Eq.\ (\ref{G0Assym}), we use
the following integral identity $\int_0^{\infty} dx xe^{2\text{Re}\beta_2(x)}sin[\pi(1-e^{-x})] = \frac{\pi}{2e^{2\bold{\gamma}}}$, where $x=t/2\pi C$ is the dimensionless variable. Introducing the bare conductance $G_R=|\tau_R|^2/2\pi v^2_{F}$, we present the final result
\begin{equation}
G_{\text{dir}}(T) = G_{R}\frac{2\pi^4T^2}{3e^{2\bold{\gamma}}E^2_C}.
\end{equation}

 Similar arguments may be used for the oscillating part of the conductance to show that it is proportional to $(T/E_C)^2$ as well. We confirm this fact by the exact numerical evaluation of the time integrals (see Fig.\ \ref{fig4.Temperature dependence of visibility in case of asymmetric barriers.}). Therefore, the final result in the case of an assymetric setup at small temperatures can be written as (\ref{Conductance.Quantum regime. Assymmetric barriers}). 

\subsection{Thermal regime}
In the limit $T\gg E_C$ the exponential factor in the integrand can be neglected and the main contribution to the integral comes from poles, i.e., from  small times $t \sim 1/\epsilon_F$.
\begin{equation}
\label{G0GR}
G_{\text{dir}}=-\frac{|\tau_R|^2}{4v^2_{F}}\int dt it \left[\frac{1}{\sinh^2(t-i0)}-\frac{1}{\sinh^2(t+i0)}\right]=\frac{|\tau_R|^2}{2\pi v^2_{F}} =G_R.
\end{equation}
Next, the oscillating term has the following form 
\begin{equation}
\label{GoscAs}
G_{\text{osc}}=-2\frac{|\tau_R|^2}{a}\exp[\alpha_{A}(T)]
{\rm Re}(\tau_L e^{-2\pi i Q_0}G_1),
\end{equation}
where
\begin{equation}
\alpha_{A}(T)=-\int \frac{dx x}{1+x^2}\frac{1}{1-e^{-x/2\pi TC}}=\bold{\gamma}+\log\left(\frac{a}{4\pi^2 v_F C}\right)+2\pi^2 TC+\log(4\pi^2 TC)+\Psi(1/4\pi^2 TC),
\end{equation}
$\gamma$ being the Euler gamma constant, $\Psi(x)$ is the digamma function and
\begin{equation}
\label{ExpG1}
G_1 = \int_{-\infty}^{\infty} dt it [K(t)A_1(t)-K^{\ast}(t)R_1(t)].
\end{equation}
Here, we introduced $K(t)=K_{\text{ff}}(t)e^{2\beta(t)}$, $K_{\text{ff}}(t)$ is given by (\ref{Kff}) and expressions for $A_1(t)$, $R_1(t)$ read
\begin{equation}
\begin{split}
& A_1(t)=\int_0^{-\infty} dt_1\left[e^{F(t-t_1)-F(-t_1)}-1\right]+\int_t^0 dt_1\left[e^{F(t-t_1)+F(t_1)}-1\right]+\int_{-\infty}^t dt_1\left[e^{-F(t_1-t)+F(t_1)}-1\right], \\
& R_1(t)=\int_t^{-\infty} dt_1\left[e^{F(t-t_1)-F(-t_1)}-1\right]+\int_0^t dt_1\left[e^{-F(t_1-t)-F(-t_1)}-1\right]+\int_{-\infty}^0 dt_1\left[e^{-F(t_1-t)+F(t_1)}-1\right], \\
\end{split}
\end{equation}
where
\begin{equation}
\label{ExpforF}
F(t_1)=4\pi iC \int d\omega \frac{1}{1+4(\pi \omega C)^2}\frac{e^{-i\omega t_1}-1}{1-e^{-\omega /T}}. 
\end{equation}

One should note, that $F^{\ast}(t_1)=-F(-t_1)$, which implies that $R_1(t)=-A^{\ast}_1(t)$,  and thus we need to calculate only the function $A_1(t)$. 
At large temperatures, $T\gg E_C$, we expand $1/(1-e^{-\omega/T}) \approx T/\omega + 1/2$ in the integrand of (\ref{ExpforF}), and after integrating  over $\omega$ we obtain
\begin{equation}
\label{ApproxF}
F(t_1)=-i\pi \left(1-e^{|-t_1/2\pi C|}\right)+4\pi^2 TC \left(1-e^{-|t_1/2\pi C|}\right)\text{sign}(t_1).
\end{equation}
 Next,  using again $T\gg E_C$ we conclude, that the main contribution to the integral (\ref{ExpG1}) comes from times smaller than $2\pi C$, because $T \gg E_C$, therefore on can expand (\ref{ApproxF}): $A_1(t) = R_1(t) \approx i(e^{2\pi Tt}-1)/T$.

Substituting  $A_1(t)$ and $R_1(t)$ into Eq.\ (\ref{ExpG1}), we arrive at the following expression
\begin{equation}
G_1=\frac{-iT}{4v^2_F C}\int dt t^2e^{-Tt^2/2C}\frac{e^{2\pi Tt}-1}{\sinh^2(\pi Tt)}, 
\end{equation}
where we have simplified $K(t)= K_{ff}(t)e^{-Tt^2/2C-it/2C}$, because the integral comes from $t/ C \ll 1$. Moreover, the main contribution comes from times $t \gg 1/T$,  so that $(e^{2\pi Tt}-1)/\sinh^{2}(\pi Tt)\approx 4$. After integrating over $t$, we get $G_1=-i\sqrt{2\pi C}/2v^2_F \sqrt{T}$.
Finally, combining this with Eqs.\ (\ref{G0GR}) and (\ref{GoscAs}), and absorbing the phase factor $e^{-i\pi/2}$ into $\tau_L$, we arrive at the result (\ref{Conductance.Thermal regime.Asymmetric barriers}) in the main text.


\end{widetext}


\begin{thebibliography}{99}

\bibitem{Grabert} Single Charge Tunneling, edited by H. Grabert and M. H. Devoret (Plenum Press, New York, 1992). 
\bibitem{Kastner} M. A. Kastner, Rev. Mod. Phys. {\bf 64}, 849 (1992).  
\bibitem{Averin} D. V. Averin and K.K Likharev, in Mesoscopic Phenomena in Solids, eds. B.L. Altshuler, P.A. Lee, and R.A. Webb (Elsevier, Amsterdam, 1991). 
\bibitem{Fulton} T. A. Fulton and G. J. Dolan, Phys. Rev. Lett. {\bf 59}, 109 (1987). 
\bibitem{Hawrylak} P. Hawrylak, L. Jacak, and A. Wojs, Quantum dots (Springer Verlag, Berlin,
1998).
\bibitem{Ando} T. Ando, Y. Arakawa, K. Furuya, S. Komiyama, and H. Nakashima, eds., Mesoscopic Physics and Electronics (Springer, 1998).
\bibitem{Larsen} T. W. Larsen, K. D. Petersson, F. Kuemmeth, T.  S. Jespersen, P. Krogstrup, J. Nyg\aa{}rd, and C. M. Marcus, Phys. Rev. Lett. {\bf 115}, 127001 (2015). 
\bibitem{Albrecht} S. M. Albrecht,	A. P. Higginbotham,	M. Madsen,	F. Kuemmeth,	T. S. Jespersen,	J. Nyg\aa{}rd,	P. Krogstrup and C. M. Marcus, Nature {\bf 531}, 206 (2016). 
\bibitem{Pierre} S. Jezouin, Z. Iftikhar, A. Anthore, F. D. Parmentier, U. Gennser, A. Cavanna, A. Ouerqhi, I. P. Levkivskyi, E. Idrisov, E. V. Sukhorukov, L. I. Glazman and F. Pierre, Nature {\bf 536}, 58 (2016). 
\bibitem{Kouwenhoven}  L. P. Kouwenhoven, N. C. van der Vaart, A. T. Johnson, W. Kool, C. J. P. M. Harmans, J. G. Williamson, A. A. M. Staring, and C. T. Foxon, Z. Phys. B {\bf 85}, 367 (1991).
\bibitem{Staring} A. A. M. Staring, J. G. Williamson, H. van Houten, C. W. J. Beenakker, L. P. Kouwenhoven,
and C. T. Foxon Physica B {\bf 175}, 226 (1991).
\bibitem{Van der Vaart} N. C. van der Vaart, A. T. Johnson, L. P. Kouwenhoven, D. J. Maas, W. de Jong, M. P. de
Ruyter van Steveninck, A. van der Enden, C. J. P. M. Harmans, and C. T. Foxon, Physica B {\bf 189}, 99 (1993).
\bibitem{Korshunov} S. E. Korshunov, JETP Lett. {\bf 45}, 434 (1987). 
\bibitem{Schon} G. Sch\"on and A. D. Zaikin, Phys. Rep. {\bf 198}, 237 (1990). 
\bibitem{Flensberg} K. Flensberg, Phys. Rev. B {\bf 48}, 11156 (1993). 
\bibitem{Nazarov} Y. V. Nazarov, Phys. Rev. Lett. {\bf 82}, 1245 (1999). 
\bibitem{Chouvaev} Chouvaev, D., Kuzmin, L. S., Golubev, D. S., and Zaikin, A. D., Phys. Rev. B, {\bf 59(16)}, 10599 (1999). 
\bibitem{Golubev} Golubev, D. S., K\" onig, J., Schoeller, H., Sch\"on, G., and  Zaikin, A. D., Phys. Rev. B, {\bf 56(24)}, 15782 (1997). 
\bibitem{Titov} M. Titov and D. B. Gutman, Phys. Rev. B {\bf 93}, 155428 (2016) 
\bibitem{Burmistrov} I. S. Burmistrov, Low Temperature Physics {\bf 43}, 95 (2017)
\bibitem{Furusaki} A. Furusaki, K. A. Matveev, Phys. Rev. B {\bf 52}, 16676 (1995). 
\bibitem{Sukhorukov1} E. V. Sukhorukov, Physica E {\bf 77}, 191 (2016). 
\bibitem{Slobodeniuk} A. O. Slobodeniuk, I. P. Levkivskyi, and E. V. Sukhorukov, Phys. Rev. B {\bf 88}, (2013). 
\bibitem{Wen} X. G. Wen, Phys. Rev. B {\bf 41}, 12838 (1990). 
\bibitem{Sukhorukov} E. V. Sukhorukov and V. V. Cheianov, Phys. Rev. Lett. {\bf 99}, 156801 (2007). 
\bibitem{Levkivskyi} I. P. Levkivskyi and E. V. Sukhorukov, Phys. Rev. B {\bf 78}, 045322 (2008). 
\bibitem{Slobodeniuk1} A. O. Slobodeniuk, E. G. Idrisov, and E. V. Sukhorukov, Phys. Rev. B {\bf 93}, 035421 (2016). 
\bibitem{Lifshitz} E. M. Lifshitz and L. P. Pitaevskii, Statistical Physics, Part 2, Landau and Lifshitz Course of Theoretical Physics Vol. 9 (ButterworthHeinemann, Oxford, 1980). 
\bibitem{Nagaosa} A. Furusaki and N. Nagaosa, Phys. Rev. B {\bf 47}, 3827 (1993) 
\bibitem{Ashoori1} R. C. Ashoori, H. L. Stormer, J. S. Weiner, L. N. Pfeiffer, S. J. Pearton, K. W. Baldwin, and K. W. West, Phys. Rev. Lett. {\bf 68}, 3088 (1992). 
\bibitem{Ashoori2} R. C. Ashoori, H. L. Stormer, J. S. Weiner, L. N. Pfeiffer, K. W. Baldwin, and K. W. West, Phys. Rev. Lett. {\bf  71}, 613 (1993).


\end{thebibliography}
\end{document}